\documentstyle[aps,prb,twocolumn,epsfig,floats,amssymb,amsfonts]{revtex}

\newcommand{\bohr}{\mu_B}

\begin{document}
\draft
\title{Magnetic-field dependence of energy levels in ultrasmall
metal grains}
\author{S. Adam, M. L. Polianski, X. Waintal$^*$, and P. W. Brouwer}
\address{Laboratory of Atomic and Solid State Physics,
Cornell University, Ithaca, NY 14853-2501\\
{\rm \today}
\medskip \\ \parbox{14cm}{\rm 
We present a theory of mesoscopic fluctuations of $g$ tensors and
avoided crossing energies in a small metal grain. The model, based on 
random matrix theory, contains both the orbital and spin contributions to
the $g$ tensor. The two contributions can be experimentally separated for
weak spin-orbit coupling while they merge in the strong coupling limit.
For intermediate coupling, substantial correlations are found between $g$
factors of neighboring levels.\medskip\\
PACS numbers: 71.70.Ej, 73.23.-b, 73.23.Hk}}
\maketitle

\section{Introduction}

\begin{figure}
\bigskip
\epsfxsize=0.8\hsize
\hspace{0.025\hsize}
\epsffile{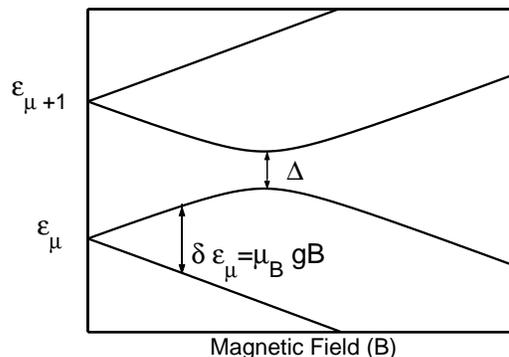}
\caption{\label{Fig:TwoLevels}
A cartoon showing the definitions of the $g$-factors and the
avoided crossing energy $\Delta$. At zero magnetic field, all
energy levels $\varepsilon_{\mu}$ are doubly degenerate. A
magnetic field splits these doublets. The 
$g$-factor measures the size of the splitting of a doublet
$\varepsilon_{\mu}$ as a function of magnetic field, see
Eq.\ (\protect\ref{eq:gfact}). The avoided crossing energy
$\Delta$ is the minimum distance at the first avoided crossing
of neighboring energy levels, see Sec.\ \protect\ref{avoid}.}
\end{figure}

Recent developments in nanofabrication techniques have allowed for 
the resolution of individual ``particle-in-a-box'' energy levels in
small metal grains or semiconductor quantum dots using tunneling 
spectroscopy.\cite{Ashoori,Kastner,Ralph,Beenakker} 
In the absence of a magnetic field, the energy levels 
$\varepsilon_{\mu}$ are two-fold
degenerate (Kramers' degeneracy). An applied magnetic field $B$ lifts
the degeneracy; the splitting of the doublet is described with the
help of a ``$g$ factor'', 
\begin{equation}
  \delta \varepsilon_{\mu} = \mu_B g B,
  \label{eq:gfact}
\end{equation}
where
$\mu_B = $ is the Bohr magneton. A cartoon of the magnetic-field
dependence of the energy levels is shown in Fig.\ \ref{Fig:TwoLevels}.
Whereas $g = 2$ for electrons in
vacuum, in a metal grain the $g$ factor can be different from two as
a result of spin-orbit scattering. Recently, the magnetic-field 
dependence of particle-in-a-box levels in metal grains have been 
measured by two groups.\cite{Salinas,Davidovic,Petta,Petta2} 
Measured
$g$ factors range from $0.1$ to $2$, depending on grain size, 
material, and, in the case of Ref.\ \onlinecite{Salinas}, doping
with heavy ions.

Unlike in bulk metals, where $g$ factors are used to describe
the effect of spin-orbit coupling on the band structure, $g$ factors
in a metal grain are not a ``bulk'' property.\cite{WPHalperin}
Not only does the
typical value of the $g$ factors depend on the size of the metal
grain, $g$ factors also depend on the microscopic details such as
the impurity configuration, the location of defects, and the form
of the grain boundary. As a result, different energy levels in a
metal grain have different $g$ factors. Moreover, even if the metal
grain is roughly spherical and without lattice anisotropy, the
presence of impurities breaks the rotational symmetry on the
microscopic scale, causing $g$ factors to depend on the direction
of the applied magnetic field. A statistical description
of the level-to-level fluctuations of $g$ factors in metal grains
has been formulated by Matveev {\em et al.}\cite{Matveev} and by
Halperin and two of the authors\cite{Brouwer}
using random matrix theory (RMT). Petta and Ralph\cite{Petta}
measured $g$ factors
for up to 9 consecutive levels in nanometer-size Cu, Ag, and Au grains 
and found good agreement with the distributions of Refs.\
\onlinecite{Matveev,Brouwer}. The dependence on the direction $\hat B$
of the magnetic field is taken into account by replacing the
$g$ factor by a ``$g$ tensor'' ${\cal G}$,\cite{Slichter} 
\begin{equation}
  \delta \varepsilon_{\mu} = \mu_B B
  (\hat B^{\rm T} {\cal G}_{\mu} \hat B)^{1/2}.
  \label{eq:gtens}
\end{equation}
(The $g$ tensor carries a subscript $\mu$ to reflect its
dependence on the energy level $\varepsilon_{\mu}$.)
The $g$-factor (\ref{eq:gfact})
for a magnetic field in the $z$ direction is the
square root of the tensor element ${\cal G}_{zz}$.
A measurement of full
$g$ tensors in Cu grains was reported quite 
recently.\cite{Petta2} Again, good agreement was found between
the experimentally measured $g$-tensor distribution and RMT.

The effect of the spin-orbit interaction on the wavefunctions in
a metal grain can be described by a dimensionless parameter
$\lambda$,
\begin{equation}
  \lambda^2 = {\pi \hbar \over \tau_{\rm so} \delta},
  \label{Eq:Lambda}
\end{equation}
where $\tau_{\rm so}$ is the spin-orbit scattering time and $\delta$
is the mean spacing between Kramers' doublets in the grain (in the
absence of the magnetic field). The effects of spin-orbit scattering
are weak if $\lambda \ll 1$. In that case, wavefunctions are real and
have a well-defined spin; the electron magnetic moment is close to its
vacuum value $g = 2$. In the opposite limit of strong spin-orbit
scattering, $\lambda \gg 1$, wavefunctions are complex and have no
well-defined spin. Hence, the spin contribution to the electron's
magnetic moment is strongly suppressed, compared to the case of
electrons in vacuum. However, in addition to a contribution from 
the electron's spin, there may be a significant orbital contribution 
to the magnetic moment carried by a single electron if spin-orbit 
scattering is present:  wavefunctions are complex, and hence 
current-carrying.\cite{Matveev}

Experimental estimates of $\lambda$ are close to zero in Al and range 
from $0.7$ in a small Cu grain ($\delta \approx 0.7\, $ meV) to $13$ 
in a larger Au grain ($\delta \approx 0.1$\, meV).\cite{Petta}  A full 
theory of
the combined orbital and spin contributions to the $g$ tensor was
developed for the asymptotes $\lambda \ll 1$ and $\lambda \gg 1$
only.\cite{Matveev,Brouwer} Both theories calculate distributions 
normalized to the average 
$(\langle g^2 \rangle)^{1/2}$. In addition, Matveev {\em et al.}
calculate both spin and orbital contributions to 
$(\langle g^2 \rangle)^{1/2}$, while Ref.\ \onlinecite{Brouwer} 
considered
the spin contribution only. The case of intermediate $\lambda$,
necessary for a quantitative comparison with the experiments of Ref.\
\onlinecite{Petta}, was studied in Ref.\ \onlinecite{Brouwer} using
numerical diagonalization of a random matrix model with variable
spin-orbit scattering strength, but without inclusion of the orbital
contribution to the magnetization.

In this paper we construct a random matrix theory that
describes both spin and orbital contributions to the electron
$g$ tensor. In the limit $\lambda \gg 1$ our model reproduces
the $g$ tensor distribution found in Refs.\
\onlinecite{Matveev,Brouwer}, but it also provides a simple model
to numerically obtain the full $g$ tensor distribution for 
arbitrary spin-orbit scattering strength. In addition to the
distribution of the $g$ tensor we also look at the correlator
of $g$ tensors of neighboring levels. While $g$ tensors are
not correlated for $\lambda = 0$ and, as we show here, 
for $\lambda \gg 1$; we find that correlations can be substantial 
for $\lambda$
of order unity. The random-matrix model is formulated in 
Sec.\ \ref{RMTsection}; the $g$ tensor distributions are
considered in Sec.\ \ref{stat}.

In addition to the $g$ factors, which describe the magnetic-field
dependence of the energy levels at very small magnetic fields, Salinas
{\em et al.} obtained additional
information on the magnitudes of spin-orbit
scattering matrix elements from avoided crossings of energy levels at
higher magnetic fields: For weak spin-orbit scattering, the minimal
energy separation $\Delta$ in an avoided crossing between the
downward moving level $\varepsilon_{\mu+1,-}$ and the upward moving
level $\varepsilon_{\mu,+}$ is twice the matrix element of the
spin-orbit coupling between the corresponding
eigenstates,\cite{Salinas} see Fig.\ \ref{Fig:TwoLevels}.
In Sec.\
\ref{avoid} we calculate the avoided crossing energy $\Delta$
from the random matrix model, and find its statistical distribution
and dependence on the direction of the magnetic field $\vec B$.



\section{Random Matrix Model}
\label{RMTsection}

In this section we formulate a random-matrix model that describes
the magnetic-field dependence of energy levels in a metal grain
with spin-orbit scattering,
taking into account both the Zeeman and the orbital effects of the
magnetic field. Following the basic premises of random matrix
theory, we replace the Hamiltonian of the metal grain by a 
$2N \times 2N$ matrix ${\cal H}$,
\begin{equation}
  {\cal H}(\lambda)  = H_{\rm GOE} + \frac{\lambda}{\sqrt{N}} 
  H_{\rm GSE} + H_{B}. \label{Eq:HSA}
\end{equation}
The first two terms on the right hand side of Eq.\ (\ref{Eq:HSA})
describe the Hamiltonian in the absence of the magnetic field;
the last term $H_B$ describes the effect of the magnetic field.
Without the magnetic field,
${\cal H}$ is taken from an ensemble that interpolates between the
Gaussian Orthogonal and Gaussian Symplectic ensembles of random
matrix theory. 
The Gaussian Orthogonal Ensemble (GOE), which is relevant for
metal grains
without spin-orbit scattering, consists of real symmetric $N \times 
N$ matrices with independently and Gaussian distributed elements,
multiplied by the $2 \times 2$ unit matrix $\openone_2$ in spin
space,
\begin{equation}
  H_{\rm GOE} = S \otimes \openone_2, \ \
  P(S) \propto e^{-(\pi^2/4 N \delta^2) {\rm tr}\, S^{\rm T} S}.
\end{equation}
Here $\delta$ is the mean level spacing in the metal grain (i.e.,
the mean spacing of the Kramers' doublets).
The Gaussian Symplectic Ensemble (GSE), which describes metal
grains with strong spin-orbit scattering, consists of self-dual
quaternion matrices.\cite{Mehta} A Hamiltonian taken from the GSE can 
be parameterized as
\begin{equation}
  H_{\rm GSE} =  {1 \over 2}
  \left( A_0 \otimes \openone_2 + i \sum_{j=1}^{3} A_j \otimes
  \sigma_j \right), \label{Eq:HGSE}
\end{equation}
where $A_0$ is a real symmetric $N \times N$ matrix and the $A_j$,
$j=1,2,3$, are real and antisymmetric $N \times N$ matrices. The
four matrices $A_0$, $A_1$, $A_2$, and $A_3$ have independently
and Gaussian distributed elements,
\begin{equation}
  P(A_{j}) \propto e^{-(\pi^2/4 N \delta^2) {\rm tr}\, A_{j}^{\rm
  T} A_{j}},\ \ j = 0,1,2,3.
  \label{Eq:PA}
\end{equation}
The crossover parameter $\lambda$ describes the strength of the 
spin-orbit scattering in the Hamiltonian of Eq.\
(\ref{Eq:HSA}). 
The cases $\lambda = 0$ and $\lambda \to \infty$ correspond to the GOE
and GSE, respectively.

The effect of the magnetic field $\vec B = (B_1, B_2, B_3)$ 
is described by the term $H_B$ in Eq.\ (\ref{Eq:HSA}),
\begin{eqnarray}\label{Eq:Bperturb}
  H_B &=& \sum_{j=1}^{3} B_j M_j,
\end{eqnarray}
where the $2 N \times 2 N$ matrices $M_j$ ($j=1,2,3$) are given
by
\begin{eqnarray}
  M_j = \mu_B \left( \openone_N \otimes \sigma_j +
    i {\pi \eta \over \delta \sqrt{N}} X_j \otimes \openone_2
  \right), 
\label{Eq:Hnorm}
\end{eqnarray}
where the $X_j$, $j=1,2,3$, are real antisymmetric matrices,
with independent and Gaussian distributions,
\begin{equation}
  P(X_{j}) \propto e^{-(\pi^2/4 N \delta^2) {\rm tr}\, X_{j}^{\rm
  T} X_{j}}.
\end{equation}
The first term in Eq.\ (\ref{Eq:Hnorm}) describes the coupling
of the magnetic field to the electron spin; the second term,
which is diagonal in spin space,
describes the coupling of the magnetic field to the orbital
angular momentum. The second term in Eq.\ (\ref{Eq:Hnorm})
was originally proposed by Pandey and Mehta to describe the
orbital effect of a time-reversal symmetry breaking magnetic 
field on the statistics of energy levels.\cite{Pandey,BeenReview}
For a diffusive spherical grain with radius
$R$, mean free path $l$, and effective electron mass $m^*$, 
the coefficient $\eta$ is given by\cite{FrahmPichard}
\begin{equation}
  \eta^2 = (m/m^*)^2 {l \over 5R}, \label{eq:alpha1}
\end{equation}
whereas for a ballistic sphere with diffuse boundary scattering, one
has\cite{erratum}
\begin{equation}
  \eta^2 = (m/m^*)^2 {1 \over 8}. \label{eq:alpha2}
\end{equation}
At the end of the calculation, the limit $N \to \infty$ is taken.
Without the orbital term, the Hamiltonian ${\cal H}$ of Eq.\
(\ref{Eq:HSA}) is the 
same as the random-matrix Hamiltonian used by Halperin and two
of the authors in Ref.\ \onlinecite{Brouwer}.

The $g$ tensor ${\cal G}$ and the avoided crossing energy $\Delta$
will be expressed in terms of matrix elements involving the
eigenvectors of the Hamiltonian (\ref{Eq:HSA}).
Eigenvectors $\psi_{\mu}$ of the Hamiltonian (\ref{Eq:HSA}) are
$2N$ component complex vectors. Their elements are denoted
as $\psi_{\mu}(n,\sigma)$, where $n=1,\ldots,N$ refers to the
``orbital''
degrees of freedom, and $\sigma = \pm 1$ to spin.
At zero magnetic field, all eigenvalues of the Hamiltonian
(\ref{Eq:HSA}) are twofold degenerate (Kramers' degeneracy): 
each eigenvalue $\varepsilon_{\mu}$ ($\mu=1,\ldots,N$) has two
orthogonal
eigenvectors $\psi_{\mu}$ and ${\cal T} \psi_{\mu}$ where 
${\cal T} \psi(n,\sigma) = \sigma\psi^*(n,-\sigma)$, is the 
time-reversed of $\psi$.
In the GOE ($\lambda = 0, B = 0$), the eigenvectors $\psi_{\mu}$ and 
${\cal T}\psi_{\mu}$ can be chosen such that $\psi_{\mu}(n,+1)
= -{\cal T}\psi_{\mu}(n,-1)$ is a real number and $\psi_{\mu}(n,-1)
= {\cal T} \psi_{\mu}(n,1) = 0$. In that case, the nonzero elements
$\psi_{\mu}(n,+1)$ 
are independently and Gaussian distributed 
with zero mean and with variance $1/N$.\cite{Mehta} 
(Of course, any
linear combination of $\psi_{\mu}$ and ${\cal T}\psi_{\mu}$ forms a 
valid pair of eigenvectors for the eigenvalue $\varepsilon_{\mu}$
as well.)
In the GSE ($\lambda \to \infty$, $B=0$), 
the elements of $\psi_{\mu}$ are
complex numbers with independent and Gaussian distributions
with variance $1/2N$. In both the GSE and the GOE different
eigenvectors are statistically uncorrelated.

In the crossover between GOE and GSE,
the eigenvector distribution
is more complicated than in each of the two basic ensembles. 
Unlike for the cases of the pure GOE and GSE, eigenvectors
at different energy levels are correlated, so that it is no longer
sufficient to look at the distribution of one eigenvector
alone.\cite{ABSW}
Since orthogonal invariance is preserved throughout the
GOE-GSE crossover, 
the problem of finding the (joint) distribution of one or more
eigenvectors in the crossover ensemble can be simplified by 
considering their orthogonal invariants first. 
For each pair of eigenstates
$\psi_{\mu}$ and $\psi_{\nu}$, the invariants are four quaternion
numbers $\rho_{\mu \nu}^{j}$, $j=0,1,2,3$.
If we diagonalize ${\cal  H}$, writing
\begin{equation}
  {\cal H}(B=0) = U (E \otimes \openone_2) U^{\dagger},
\end{equation}
where $U$ is the symplectic eigenvector matrix and
the $N \times N$ diagonal matrix $E$ contains the eigenvalues
$\varepsilon_{\mu}$ on the diagonal, they are
\begin{eqnarray}
  \rho^0_{\mu \nu} &=& [U^{\dagger} U]_{\mu \nu} 
  \nonumber \\
  &=& \delta_{\mu \nu} \openone_{2} \\
  \rho^j_{\mu \nu} &=& i [U^{\dagger} \sigma_j U]_{\mu \nu}
  \nonumber \\ &=&
  \left( \begin{array}{ll} (\rho_{\mu \nu}^{j})_{++}
  & (\rho_{\mu \nu}^{j})_{+-} \\
  (\rho_{\mu \nu}^{j})_{-+}
  & (\rho_{\mu \nu}^{j})_{--} \end{array} \right),\ \ j=1,2,3.
\end{eqnarray}
The $\rho_{\mu \nu}^{j}$ satisfy a criterion of anti-hermiticity,
\begin{equation}
  \rho_{\mu\nu}^{j} = -(\rho_{\nu\mu}^{j})^{\dagger},\ \ j=1,2,3.
\end{equation}
The orthogonal invariants $\rho_{\mu\nu}^0$ express orthonormality of
the
eigenvectors $\psi_{\mu}$ and ${\cal T}\psi_{\mu}$. The remaining
orthogonal invariants $\rho_{\mu\nu}^j$ 
are characteristic for the crossover and
determine to what extend spin-rotation symmetry has been broken.  
In the GOE, we have: $\sum_{ k} {\rm tr}(\rho_{\mu\mu}^{i}\sigma_{k}) 
{\rm tr} (\rho_{\mu\mu}^{j}\sigma_{k})
= 4\delta_{ij}$,   
while $\rho_{\mu\nu}^j = 0$ if $\mu \neq \nu$; in the GSE, $\rho_{\mu
\nu}^{j} = 0$ for all $\mu$ and $\nu$.
An average involving different eigenvectors is then calculated in two
steps: First, eigenvector elements 
have a Gaussian distribution with zero mean and with variance 
determined by the orthogonal invariants.\cite{ABSW} 
In spinor notation,
where $\psi(n)$ denotes the 2-component spinor with elements
$\psi(n,+1)$ and $\psi(n,-1)$, these variances are
\begin{eqnarray}
  \langle \psi_{\mu}(n)^{\dagger} \psi_{\nu}(m) 
  \rangle &=& {\delta_{mn} \over N} \delta_{\mu \nu}, \nonumber \\
  i \langle \psi_{\mu}(n)^{\dagger} \sigma_j
  \psi_{\nu}(m) \rangle &=& {\delta_{mn} \over N}
  (\rho_{\mu\nu}^{j})_{++}, 
  \nonumber \\ 
  \langle \psi_{\mu}(n)^{\rm T} \sigma_2\psi_{\nu}(n)
  \rangle &=& 0, \nonumber \\
  \langle \psi_{\mu}(n)^{\rm T} \sigma_2 \sigma_j \psi_{\nu}(n)
  \rangle &=& {\delta_{mn} \over N}
  (\rho_{\mu\nu}^{j})_{-+}.
  \label{eq:rhoavg}
\end{eqnarray}
With the help of Eq.\ (\ref{eq:rhoavg}) any average over eigenvectors
can be expressed in terms of the orthogonal invariants involved in
the problem. 

What remains is to find the average over a small number
of orthogonal invariants.
To our knowledge, a solution of this problem exists for the
limits $\lambda \ll 1$ and $\lambda \gg 1$ only. 
For strong spin-orbit scattering, $\lambda \gg 1$, the solution 
takes the form of a surmise
equating the distribution of the $\rho_{\mu\nu}^{j}$ for the 
$2N \times 2N$ crossover Hamiltonian (\ref{Eq:HSA}) to the 
distribution of the same quantities for a GSE Hamiltonian of
a smaller size $2N'$,\cite{ABSW} 
\begin{eqnarray}
  N' &=& \lambda^2 N (\lambda^2 + 2N)/(\lambda^2 + N)^2
  \nonumber \\ &\to& 2 \lambda^2 \ \ \mbox{if $N \to \infty$},
\label{Eq:MN}
\end{eqnarray} 
provided the energy difference $|\varepsilon_{\mu} -
\varepsilon_{\nu}| \ll \lambda^2 \delta$.
This means that the elements of the matrix $\rho^{j}$ are
uncorrelated and that they have a Gaussian distribution with 
variance
\begin{eqnarray}
  \langle |(\rho_{\mu \nu}^{j})_{++}|^2 \rangle &=& \frac{1}{2N}, 
		\nonumber \\
 \langle |(\rho_{\mu \nu}^{j})_{+-}|^2 \rangle &=& 
		\frac{1+\delta_{\mu\nu}}{2N}.
\end{eqnarray}

A similar surmise was proposed in Ref.\ \onlinecite{ABSW}
for the eigenvector statistics in the crossover between the GOE 
and the Gaussian Unitary Ensemble of random-matrix theory.
The motivation of this surmise becomes clear once 
we consider the crossover Hamiltonian (\ref{Eq:HSA})
in the eigenvector basis of $H_{\rm GOE}$.\cite{ABSW} 
In this basis, eigenvectors of the crossover Hamiltonian 
are ``localized'': they are mainly built up from eigenvectors of 
$H_{\rm GOE}$ with energies inside a window of size 
$\sim N' \delta$ (with $N'$ to be determined later). Since 
changing to the GOE basis does not change orthogonal invariants, 
we can calculate the $\rho_{\mu\nu}^{j}$
using an effective $2N' \times 2N'$ Hamiltonian that contains 
the $2N'$ relevant GOE eigenvectors only, if $|\mu - \nu| \ll N'$. As
the spin-rotational symmetry breaking term is large for the
effective Hamiltonian, its distribution is that 
of the GSE, not a crossover. The exact
relation (\ref{Eq:MN}) between $N'$ and $N$ is found
matching the distributions of a single orthogonal invariant
$\rho_{\mu\mu}^{j}$ in the crossover Hamiltonian and in the
GSE.\cite{Brouwer}

In the following two sections, 
the random matrix model (\ref{Eq:HSA}) will serve as a starting
point for analytical calculations of the $g$ tensor distribution
and avoided crossing energies in the regimes of weak
spin-orbit scattering, $\lambda \ll 1$, and of strong spin
orbit scattering, $\lambda \gg 1$, and for numerical calculations
of the $g$-tensor distribution in the crossover regime $\lambda
\approx 1$. The case of weak spin-orbit scattering can be treated
using perturbation theory in $\lambda$; for strong spin-orbit
scattering, we use the full eigenvector distribution of 
the GOE-GSE crossover Hamiltonian and the surmise for the
orthogonal invariants that was discussed in this section.


\section{Statistics of the $g$ tensor}\label{stat}

A typical plot of the magnetic field dependence of energy levels
is shown in Fig.\ \ref{Fig:TwoLevels}. 
A magnetic field $\vec B = B \hat B$ splits the Kramers' doublets
$\varepsilon_{\mu}$ into pairs $\varepsilon_{\mu,\pm}$ that
depend linearly on the magnitude $B$ of the magnetic field,
\begin{equation}
  \varepsilon_{\mu,\pm} = \varepsilon_{\mu} \pm {1 \over 2}
  \delta \varepsilon_{\mu}, 
  \label{eq:GB}
\end{equation}
with $\delta \varepsilon_{\mu}$ expressed in terms of the
$g$ tensor ${\cal G}_{\mu}$ as in Eq.\ (\ref{eq:gtens}) above.
%
%
Following Ref.\ \onlinecite{Brouwer}, the $g$ tensor can be
written as
\begin{mathletters} \label{Eq:GGG}
\begin{eqnarray}
  {\cal G} = G^{\rm T} G,
\end{eqnarray}
where the $3 \times 3$ matrix $G$ has elements
\begin{eqnarray}
   G_{1j} &=&
  {2 \over \mu_B} \mbox{Re}\,
  \langle \psi_{\mu} | M_j | {\cal T} \psi_{\mu} \rangle ,
  \nonumber \\
  G_{2j} &=&
  {2 \over \mu_B} \mbox{Im}\,
  \langle \psi_{\mu} | M_j | {\cal T} \psi_{\mu} \rangle,
  \\
 G_{3j} &=&
  {2 \over \mu_B} \langle \psi_{\mu} | M_j | \psi_{\mu} \rangle ,
  \nonumber 
\end{eqnarray}
\end{mathletters}%
where $M_j$ is defined in Eq.\ (\ref{Eq:Hnorm}), $\psi_{\mu}$
is an eigenvector of ${\cal H}$ at $B=0$ with eigenvalue
$\varepsilon_{\mu}$, and ${\cal T} \psi_{\mu}$ is its
time-reversed.

The tensor ${\cal G}$ has three eigenvectors and
 three eigenvalues $g_j^2$, $j=1,2,3$. The eigenvectors are
referred to as ``principal axes'', the eigenvalues $g_1$,
$g_2$, and $g_3$ as ``principal $g$-factors''. The three 
principal $g$ factors describe the splittings of 
the doublet for magnetic fields along each of the three
principal axes.
We describe the distribution of the $g$ tensor in terms of
the distributions of its eigenvectors (the principal axes)
and eigenvalues (the principal $g$-factors).
For a roughly spherical grain, the principal axes will
be oriented randomly in space. Hence, it remains to find the
distribution of the three principal $g$ factors $g_{\mu,1}$,
$g_{\mu,2}$, and $g_{\mu,3}$. We will now consider the cases
of weak and strong spin-orbit scattering separately.
\subsection{Weak spin-orbit scattering}

To leading order in $\lambda \ll 1,\eta\lambda\ll 1$ the 
$g$ tensor reads
\begin{eqnarray}\label{Eq:gnew}
  {\cal G}_{\mu;ij} &=&
  4\left(\delta_{ij} +
  { \eta \lambda \pi \over N \delta}
  \sum_{\nu \neq \mu}
  {X^{\mu \nu}_{i} A^{\mu \nu}_{j}
  + A^{\mu \nu}_{i} X^{\mu \nu}_{j} 
  \over \varepsilon_{\mu} - \varepsilon_{\nu}}
\right.\nonumber \\ && \mbox{} \left.
  - { \lambda^2 \over N}
  \sum_{\nu \neq \mu}
  {  
  \delta_{ij} \sum_{k=1}^{3}(A^{\mu \nu}_{k})^2-
A^{\mu \nu}_{i} A^{\mu \nu}_{j}
  \over
  (\varepsilon_{\mu} - \varepsilon_{\nu})^2}\right).
  \label{Eq:Gperturb}
\end{eqnarray}
Here  $\varepsilon_{\mu}$ and  $\varepsilon_{\nu}$ are  eigenvalues of
the  Hamiltonian (\ref{Eq:HSA})  at  zero magnetic  field and  without
spin-orbit scattering,  and $A^{\mu \nu}_j$ and  $X^{\mu \nu}_{j}$ are
the  matrix  element of  the  matrices  $A_j$  and $X_j$  between  the
corresponding        eigenvectors       $|\psi_{\mu}\rangle$       and
$|\psi_{\nu}\rangle$   of  ${\cal   H}$,   respectively,  cf.\   Eqs.\
(\ref{Eq:HGSE}) and  (\ref{Eq:Hnorm}). In Eq.\  (\ref{Eq:Gperturb}) we
neglected terms  of order $\lambda^2  \eta,(\lambda\eta)^2$, which
are  small compared  to $\lambda  \eta$, and  of third  (and higher)
order  in  $\lambda$.  The  second  term  in Eq.\  (\ref{Eq:Gperturb})
corresponds  to orbital paramagnetism,  and is  of  order $\eta
\lambda$  because the  orbital  contribution couples to  complex
wavefunctions that  occur for $\lambda \neq 0$.   The $\lambda^2$ term
is a reduction of the Pauli paramagnetism caused by interaction with 
other energy levels and for the case of $i=j=3$, agrees with earlier 
work.\cite{Sone}

The distribution of ${\cal G}$ without the orbital contribution
(second term in Eq.\ (\ref{Eq:Gperturb})) was studied in Refs.\
\onlinecite{Matveev} and \onlinecite{Brouwer}. 
For very small spin-orbit scattering, however,
the orbital part is found to dominate the $g$ tensor fluctuations,
since it is of first order in the spin-orbit scattering strength
$\lambda$, while the Zeeman contribution is of order $\lambda^2$. 
Whereas the Zeeman contribution always gives $g$ factors
smaller than two --- the last term in Eq.\ (\ref{Eq:Gperturb}) is
negative definite --- the orbital contribution can be of arbitrary
sign, allowing for principal $g$ factors larger than two.
To illustrate this feature, we calculate the tails of the joint 
distribution $P(g_{1},g_{2},g_{3})$ of the three principal $g$
factors. The distribution of the tails
is dominated by events where the spacing between the level 
$\varepsilon_{\mu}$ and one of its neighbors $\varepsilon_{\mu+1}$ or 
$\varepsilon_{\mu-1}$ is exceptionally small, of order 
$\lambda \delta$ or $\lambda\eta \delta$ (whichever is larger). 
Hence, the tails of $P(g_{1},g_{2},g_{3})$
can be calculated limiting attention to the nearest-neighbor
terms in the summations in Eq.\ (\ref{Eq:Gperturb}). We order
the three principal $g$ factors as $g_1 < g_2 < g_3$ and 
parameterize them as $g_j = 2(1+y_j)$, $j=1,2,3$. 
Hence, the tails
of the distribution correspond $|y_j| \gg \max(\lambda^2,
\lambda \eta)$ for at least one of the $y_j$. 
With the definition $\Theta(x) = 1$ for $x > 0$, and $\Theta(x) = 0$ for
$x < 0$, the tails of the distribution
are found to be
\begin{eqnarray}\label{Eq:Pdistrib}
&&P(y_1<y_2<y_3)=\frac {3(\pi \lambda)^2}{8\eta^3}\Theta(-y_2)\nonumber \\ 
&&\times\frac {y_3-y_1}
{(-\pi y_2)^{7/2}}\exp\left[\frac{(y_1-y_2
-y_3)^2+4(y_1-y_2)y_2}{4\eta^2y_2}\right].
\end{eqnarray}  
In the limit $\eta \ll \lambda$ the tail of the distribution factors as
\begin{eqnarray}\label{Eq:alphasmall}
  P( y_1< y_2< y_3)&\propto &\frac {y_3-y_1}{\eta^3(-y_2)^{7/2}}
\exp\left[\frac{y_1- y_2}{\eta^2}+\frac{y_3^2}{4\eta^2 y_2}\right],
\nonumber
\end{eqnarray}
reproducing the result 
\begin{eqnarray}
P = \frac{3\lambda^2}{4\pi y_2^2}
\delta( y_3)\delta( y_2- y_1)
\end{eqnarray} 
for the tail of
the $g$ tensor distribution obtained in
Ref.\ \onlinecite{Brouwer} in the limit $\eta \to 0$. 
In the opposite limit $\lambda\ll\eta$, Eq.\ (\ref{Eq:Pdistrib})
simplifies to
\begin{eqnarray}\label{Eq:alphabig}
P(y_1,y_2,y_3)& =& 
\frac {9\eta^2\lambda^2\Theta(-y_1)\delta(y_2)\Theta(y_3)}{\pi(y_1-y_3)^4}
.
\end{eqnarray}
Equation (\ref{Eq:alphasmall}) is valid if $|y_1|, |y_2| \gg \lambda^2$;
Eq.\ (\ref{Eq:alphabig}) holds if $|y_1|, |y_3| \gg \lambda \eta$.

\begin{figure}
\epsfxsize=0.95\hsize
\hspace{0.025\hsize}
\epsffile{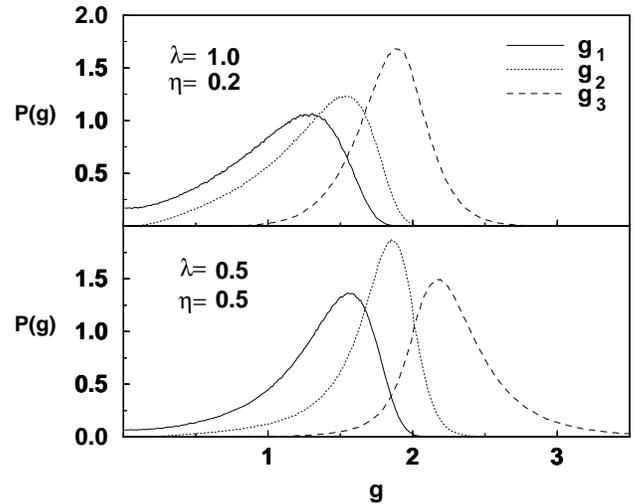}
\caption{\label{fig:Gweak} Distributions of magnitudes of the 
principal $g$ factors. Upper panel: $\lambda=1.0,\ \eta=0.2$; 
Lower panel: $\lambda=\eta= 0.5$. 
}\end{figure}

In Fig.\ \ref{fig:Gweak} we have shown the distributions of the
principal $g$ factors $g_1$, $g_2$, and $g_3$, calculated from 
the random matrix model (\ref{Eq:HSA}) using numerical diagonalization. 
Although the limits (\ref{Eq:alphasmall}) and (\ref{Eq:alphabig})
were derived for the tail of the $g$-tensor distribution only, 
they can
account for some qualitative features of the full $g$-tensor 
distribution for weak spin-orbit scattering shown in Fig.\
\ref{fig:Gweak}: when the orbital contribution to the 
$g$ tensor dominates ($\eta \gg \lambda$), generically 
$g_3 > 2$, $g_2 \approx 2$, and $g_1 < 2$, cf.\ Eq.\ 
(\ref{Eq:alphabig}). On the other hand, when the 
Zeeman contribution to the $g$ tensor dominates ($\eta \ll 
\lambda$), one typically has $g_1 \approx g_2 < 2$ and $g_3 \approx
2$, cf.\ Eq.\ (\ref{Eq:alphasmall}).

We now turn our attention to correlations between $g$ tensors of
neighboring levels. Such correlations are described by the 
correlator
\begin{eqnarray}
{\cal C}_{ij,kl}=
  \langle g_{\mu}^2 \rangle^{-2}
  \left(
  \langle{\cal G}_{\mu,ij}{\cal G}_{\mu+1,kl}
\rangle-\langle{\cal G}_{\mu,ij}\rangle
\langle{\cal G}_{\mu+1,kl}\rangle \right).
  \label{eq:C}
\end{eqnarray}
Calculating the correlator ${\cal C}$ to leading order in
$\lambda \ll 1$, we find that the result is dominated by 
events where the levels $\varepsilon_{\mu}$
and $\varepsilon_{\mu+1}$ are very close. Since this contribution
is formally divergent, as a result of the presence of the energy
denominators in the perturbation expression (\ref{Eq:Gperturb}),
a cut off must be imposed at energy separations
$\varepsilon_{\mu+1}-\varepsilon_{\mu}$ of order $\lambda \delta$
where the perturbation theory is not valid.
To treat the contribution from nearby levels $\varepsilon_{\mu+1}$
and $\varepsilon_{\mu}$ correctly, we calculate the contribution
from such events non-perturbatively. To leading
order in $\lambda \ll 1$, the result of such a treatment amounts to the
replacement of the energy denominator $\varepsilon_{\mu+1}-
\varepsilon_{\mu}$ in Eq.\ (\ref{Eq:Gperturb}) by
$[(\varepsilon_{\mu+1}-\varepsilon_{\mu})^2 +
|\vec A|^2\lambda^2/N]^{1/2}$, where $\vec A$ is shorthand 
notation for the vector with components $A_j^{\mu,\mu+1}$, $j=1,2,3$.  
We then obtain the following result:
\begin{eqnarray}
{\cal C}_{ijkl} &=& \frac{\lambda^2}{\pi}(\delta_{ik}
\delta_{jl}+\delta_{il}\delta_{jk})(\eta^2\ln\lambda+\frac{1}{20}) \nonumber \\
&+& \frac{3\lambda^2}{10\pi}\delta_{ij}\delta_{kl}. \label{Eq:Ccorr}
\end{eqnarray}

\begin{figure}
\epsfxsize=0.8\hsize
\hspace{0.1\hsize}
\epsffile{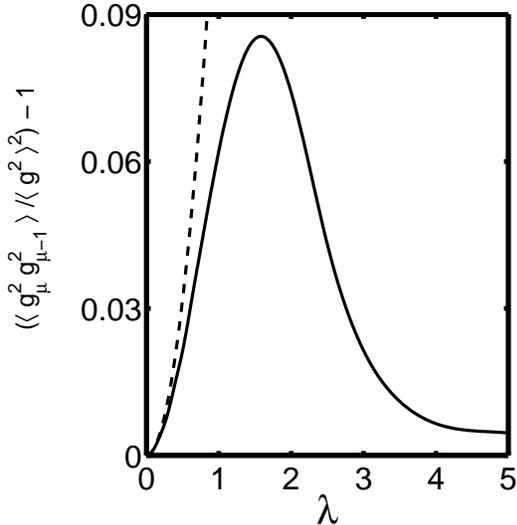}
\caption{\label{Fig:2} 
 $g$-factor correlation  as a function of  spin-orbit coupling $\lambda$
 computed   numerically  for  $200   \times  200$   GOE-GSE  crossover
 matrices.  Dashed  line shows  the  result  from perturbation  theory
 Eq.~(\protect\ref{Eq:GcorPT}).}
\end{figure}

The correlator between $g$ factors 
(at a fixed direction of the magnetic field) is found from 
Eq.\ (\ref{Eq:Ccorr}) setting $i=j=k=l=\hat B$ in the direction 
of magnetic field,
\begin{eqnarray}
  {\cal C}&=& 
  \langle g^2_{\mu+1} g^2_{\mu} \rangle/\langle g^2 \rangle^2 -
  1
=\frac{2 \lambda^2}{\pi}\left(\eta^2\ln\lambda+\frac 15\right).
  \label{Eq:GcorPT}
\end{eqnarray}

\subsection{Strong spin-orbit scattering}\label{strong}

In the regime of a strong spin-orbit scattering, $\lambda \gg 1$, 
the $g$ tensor distribution can be calculated from Eq.\ (\ref{Eq:GGG})
using the known distribution of the eigenvectors of the random
Hamiltonian (\ref{Eq:HSA}) at zero magnetic field, see Sec.\
\ref{RMTsection}. We then find that the matrix elements
of the $3 \times 3$ matrix $G$ of Eq.\ (\ref{Eq:GGG}) are Gaussian
random numbers, with zero mean and with variance $1/\lambda^2
+ 2 \eta^2$.  From this we conclude that the distribution
of the principal $g$ factors is \cite{Brouwer}
\begin{equation}
  P(g_1,g_2,g_3) \propto \left(\prod_{i<j} |g_i^2 - g_j^2|\right) \prod_{i} 
  e^{-3 g_i^2/2 \langle g^2 \rangle},  \label{eq:PgGSE0}
\end{equation}
where 
\begin{equation}
  \langle g^2 \rangle = 
  {1 \over 3} \langle g_1^2 + g_2^2 + g_3^2 \rangle
  = {3 \over \lambda^2} + {6 \eta^2}.
  \label{Eq:gsquared}
\end{equation}
Values for $\eta$ for diffusive and ballistic spherical grains
are given in Eqs.\ (\ref{eq:alpha1}) and (\ref{eq:alpha2}).
Equations (\ref{eq:PgGSE0}) and (\ref{Eq:gsquared}) extend the
result of Ref.\ \onlinecite{Brouwer} to the case $\eta\neq 0$.
Equation (\ref{Eq:gsquared}), which was derived using the random
matrix model (\ref{Eq:HSA}), agrees with the results of Matveev
{\em et al.}, which were derived using a comparison of the $g$
factors and the energy absorption of a time-dependent magnetic 
field.

\begin{figure}
\bigskip
\epsfxsize=1.\hsize
\epsffile{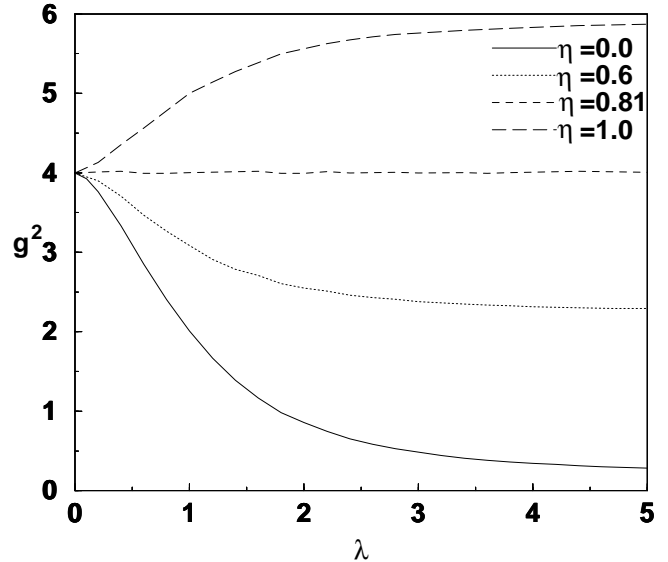}
\caption{\label{Fig:crossover} 
 Averaged $|{\vec g}|^2$ as a function of spin-orbit 
strength $\lambda$.
The critical value $\eta_0=\sqrt{2/3}\approx 0.81$.}
\end{figure}

In Fig.\ \ref{Fig:crossover} we show the result of numerical
calculations of $\langle g^2 \rangle$ as a function of the
spin-orbit scattering rate $\lambda$ and for various values of
$\eta$. For $\eta^2 < 2/3$, $\langle g^2 \rangle < 2$ for
all $\lambda$, while for $\eta^2 > 2/3$, $\langle g^2 \rangle
> 2$. The derivatives with $\lambda$ are maximal near $\lambda=0$
because of the enhanced fluctuations due to the orbital part at
small $\lambda$, cf.\ Eq.\ (\ref{Eq:Gperturb}).

Correlations between $g$ tensors of neighboring levels 
trivially vanish for large $\lambda$ because, in the GSE, 
different eigenvectors are statistically uncorrelated. 
However, since the average $g$ tensor also depends on $\lambda$, 
it is a more meaningful question to study the correlator between
$g$ tensors, normalized by the average $g$ factor, cf.\ Eq.\
(\ref{eq:C}).
In the presence of an orbital contribution
to the $g$ tensor, the average $g$ factors are nonzero for
$\lambda \gg 1$, see Eq.\ (\ref{Eq:gsquared}), so
that the vanishing of correlations in the GSE implies that they 
vanish compared to the average as well.
Without the orbital contribution,
$g$-tensor correlations cannot be addressed with 
reference to the eigenvector statistics in the GSE, because
${\cal G} = 0$ in the GSE. Instead we need the more detailed
knowledge of the eigenvector distribution for large $\lambda$,
which is summarized in Sec.\ \ref{RMTsection}. The main result 
of that section is
that the eigenvector distribution depends on the 
distribution of certain orthogonal invariants $\rho_{\mu\nu}^{j}$,
$j=1,2,3$ which are $2 \times 2$ matrices in spin space,
see.\ Eq.\ (\ref{eq:rhoavg}). With the help of Eq.\ (\ref{Eq:GGG}),
one easily verifies that, 
in the case $\eta=0$, the 
$g$ tensor may be expressed in terms of these orthogonal invariants
only,
\begin{equation}
  ({\cal G}_{\mu})_{ij} = 2
  \mbox{tr}\, \rho_{\mu\mu}^{i} \rho_{\mu\mu}^{j},\ \
  i,j=1,2,3,
\end{equation}
where the trace is taken in spin space.
Since, for $\lambda \gg 1$, the orthogonal invariants
$\rho_{\mu\mu}^j$ are all independently distributed for different
levels, we conclude that $g$ tensors of different levels are 
uncorrelated in the case $\eta=0$ as well. 

Figure \ref{Fig:2} shows the $g$-factor correlator (\ref{Eq:GcorPT})
normalized by the average $g$ factor as a function of $\lambda$.
The numerical diagonalization confirms our previous conclusions
that $g$ factor correlations are small for both asymptotic regimes
$\lambda \ll 1$ and $\lambda \gg 1$. Correlations are maximal 
for intermediate spin-orbit scattering strengths, $\lambda \sim 1.5$,
but never amount to more than $10\%$ of the average $\langle g^2 \rangle$.



\section{Avoided Crossing Energies}\label{avoid}

Once the Kramers' doublets are split by the magnetic field, half
of the levels move upward with slope $\sim (1/2) g \mu_B B$,
while the other half moves downward with the same slope. Hence,
a downward moving level $\varepsilon_{\mu+1,-}$ and the upward
moving level $\varepsilon_{\mu,+}$ meet at magnetic field strength
\begin{equation}
  B_{\rm c} = {2 (\varepsilon_{\mu+1} - \varepsilon_{\mu}) \over
  \mu_B(g_{\mu} + g_{\mu+1})}.
\end{equation}
In fact, since the matrix element
of the coupling $H_B$ to the magnetic field between the
corresponding eigenstates $|\psi_{\mu+1,-}\rangle$ and 
$|\psi_{\mu,+}\rangle$ is finite, the two levels do not cross,
but exhibit an avoided crossing, see Fig.\ \ref{Fig:TwoLevels}.
In this section we calculate the minimum distance $\Delta$ between the
energy levels in the avoided crossing, its dependence on
the direction $\hat B$ of the magnetic field, and its level-to-level
fluctuations. 

The avoided crossing energy is well-defined only
if the magnetic field dependence of the two levels
$\varepsilon_{\mu+1,-}(B)$ and $\varepsilon_{\mu,+}(B)$ is
linear, the only exception being the curvature resulting from their
mutual interaction at the avoided crossing. For the magnetic
field strengths of interest, $B \sim B_{\rm c}$, other sources of 
level curvature as a function of the magnetic field, which arise
both from the spin and orbital couplings in the Hamiltonian $H_B$
of Eq.\ (\ref{Eq:Bperturb}), are small if both $\lambda \ll 1$ 
and $\eta \ll 1$.
Hence, for the purpose of calculating the avoided crossing energy
$\Delta$ it is sufficient to consider the perturbative regime of small
$\lambda$ and small $\eta$.

Considering the Hamiltonian in the basis of states 
$|\psi_{\mu+1,-}\rangle$ and
$|\psi_{\mu,+}\rangle$, corresponding to the energy levels
$\varepsilon_{\mu+1,-}$ and $\varepsilon_{\mu,+}$ at zero magnetic
field, respectively, 
\begin{eqnarray}
  {\mathcal H} &=& 
  \left( \begin{array}{cc} \varepsilon_{\mu+1} - 
     \frac 12 \bohr Bg_{\mu} &  
  \langle \psi_{\mu+1,-} |{\cal H}_{B}| 
  \psi_{\mu,+} \rangle \\  
  \langle \psi_{\mu,+} |{\cal H}_{B}| \psi_{\mu+1,-} \rangle   &
  \varepsilon_{\mu}+\frac 12\bohr B g_{\mu-1}
  \end{array} \right), 
\end{eqnarray} 
we find that the avoided crossing energy $\Delta$ reads
\begin{eqnarray}
\Delta &=& 2 |\langle \psi_{\mu+1,-} |{\cal H}_{B_{\rm c}}| 
  \psi_{\mu,+} \rangle| \nonumber \\
  &=& {4 |\varepsilon_{\mu+1}-\varepsilon_{\mu}|
  \over \mu_B (g_{\mu+1} + g_{\mu})}
  |\langle \psi_{\mu+1,-}|\hat B \cdot \vec M| \psi_{\mu,+} \rangle|.
  \label{Eq:Delta}
\end{eqnarray}
Using first order perturbation theory in $\lambda$ and $\eta$,
we find
\begin{eqnarray}
\Delta &=& \lambda\left|\hat B\times\left(\frac{1}{\sqrt{N}}
  \vec A^{\mu+1,\mu} \right) \right|,
\end{eqnarray}
plus terms of order $\lambda \eta$ which are not relevant in the
regime we consider. The components
of the vector $\vec A^{\mu+1,\mu}$ are matrix elements
of the spin-orbit matrices $A_{j}$, $j=1,2,3$ of Eq.\ (\ref{Eq:HGSE})
in the basis that diagonalizes the Hamiltonian to zeroth order 
in $\lambda$.

In order to find the distribution
of the avoided crossing energy $\Delta$, we write 
\begin{equation}
 \Delta  = \Delta_{0} \sin \theta, \label{Eq:DeltaParam}
\end{equation}
where $0 \le \theta \le \pi $ is the angle between the direction $\hat B$
of the applied magnetic field and the vector ${\vec A}^{\mu+1,\mu}$.
Using the known distribution (\ref{Eq:PA})
of the spin-orbit coupling matrices
$A_j$ ($j=1,2,3$), one finds that the three elements
of $\vec A^{\mu+1,\mu}$ each have a Gaussian 
distribution with zero mean and with variance $N \delta^2/\pi^2$.
Hence, we conclude that the vector ${\vec A}^{\mu+1,\mu}$
is randomly oriented in space, so that
\begin{equation}
  P(\theta) = {1 \over 2} \sin \theta,
\end{equation}
and that
\begin{eqnarray}
P(\Delta_0)= 
	\frac{(\Delta_0 \pi)^2 \sqrt{2 \pi}}{(\lambda\delta)^3}
  \exp\left[-\frac 12\left(\frac{\pi\Delta_0}{\lambda\delta}\right)^2
  \right].
  \label{Eq:PDelta}
\end{eqnarray}
Equations (\ref{Eq:DeltaParam})--(\ref{Eq:PDelta}) not only give the
full distribution of the avoided crossing energy $\Delta$, but also
the dependence of $\Delta$ on the direction $\hat B$ of the magnetic
field. Equations
(\ref{Eq:DeltaParam})--(\ref{Eq:PDelta}) can be combined to give
\begin{eqnarray}
\label{Eq:IsoResultsA}
P(\Delta)=
\frac{\pi^2 \Delta}{(\lambda\delta)^2}
\exp\left[-\frac 12\left(\frac{\pi\Delta}
{\lambda\delta}\right)^2\right].
\end{eqnarray}
The latter result is relevant for comparison with experiments where
the direction of the magnetic field cannot be 
varied.\cite{Salinas,Petta}

\begin{figure}
\bigskip
\epsfxsize=0.95\hsize
\hspace{0.025\hsize}
\epsffile{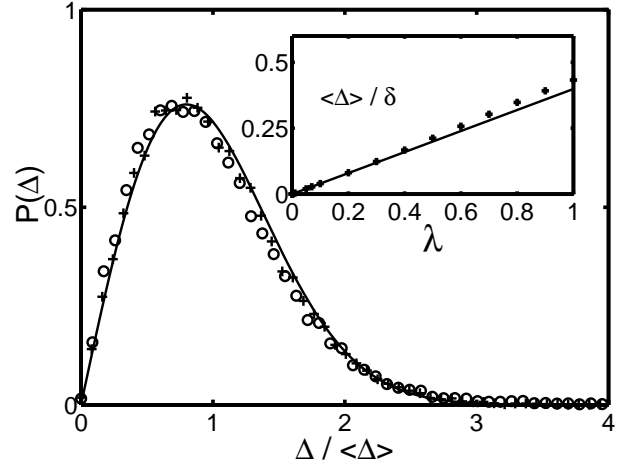}
\caption{\label{Fig:1} 
Main panel: Distribution of the avoided crossing energy
$\Delta$. 
Solid line is the perturbative result (\protect\ref{Eq:IsoResultsA});
the data points are from numerical evaluation of 
Eq.~(\protect\ref{Eq:Delta}) using the numerical diagonalization
of the random matrix model
(\protect\ref{Eq:HSA}) with $\eta = 0$ and $\lambda  =  0.2$
 (crosses), and  $\lambda  =  0.4$ (circles). 
Inset:  Comparison of perturbation theory (solid curve) and numerical 
results (data points)
for the average $\langle \Delta \rangle$.}
\end{figure}

Figure~\ref{Fig:1}  shows the distribution (\ref{Eq:IsoResultsA}),
together with results from a numerical calculation of the
distribution of
Eq.~(\ref{Eq:Delta}) using the random matrix model (\ref{Eq:HSA})
for $\eta=0$ and two different values of $\lambda$. We see that
the agreement between the numerical diagonalization of the random
matrix model and the distribution
(\ref{Eq:IsoResultsA}) calculated using first order perturbation
theory in $\lambda$ remains good up to $\lambda \sim 1$. [We should
note, however, that the approximations leading to an avoided
crossing energy that is dominated by matrix elements involving
two neighboring levels only, is valid for $\lambda \ll 1$ and
$\eta \ll 1$ only, see the discussion preceding Eq.\ 
(\ref{Eq:Delta}).] Although
there are corrections to $P(\Delta)$ to second order in $\lambda$,
the first nonzero corrections to the average $\langle \Delta 
\rangle$ appear to third order in $\lambda$ 
only.

\section{Discussion and conclusions}\label{discussion}

In this paper, we have presented a random matrix theory for the
distributions of $g$ tensors and avoided crossing energies in
small metal grains with spin-orbit scattering. Our theory 
includes both the spin and the orbital effects of the magnetic
field. 

For large spin-orbit scattering, the main effect of the 
orbital contribution is to increase the typical size of the 
$g$ tensor; the fluctuations (normalized by the average) and
the relative magnitudes of the three principal $g$ values
are the same with and without a large orbital 
contribution.\cite{Brouwer}
For weak spin-orbit scattering, the presence of an orbital
contribution to the $g$ tensor not only increases the average
of the $g$-tensor distribution,
it also changes the relative magnitudes of the principal
$g$ values. Without orbital contribution, two
principal $g$ values are approximately equal and smaller
than two, while the third principal $g$ value is close to 2.
If the orbital contribution is large, all three principal
$g$ values are different and, on average, symmetrically
positioned around two.

Petta and Ralph have measured distributions of $g$ factors 
(i.e., the square root of the ${\cal G}_{zz}$ element of the 
$g$-tensor) for 
small particles of different metals and found that 
distributions, if normalized to the average, were in very
good agreement
with the random matrix theory of Ref.\ \onlinecite{Brouwer}.
The average of the distribution, however, was up to a factor 10
smaller than the theoretical prediction (\ref{Eq:gsquared}) 
with a reasonable estimate for the parameter 
$\eta$.\cite{Matveev} 
A similar discrepancy between a experimental and
theoretical estimates
was reported in a different context
by Marcus {\em et al.}\cite{Marcus1} for the magnetic
field scale for fluctuations of Coulomb blockade heights
in two-dimensional $\mu$m-size
GaAs/GaAlAs quantum dots (see also
Ref.\ \onlinecite{Marcus2}).
Although the experimental system studied in Refs.\ 
\onlinecite{Marcus1,Marcus2} is quite different from that of
Petta and Ralph, the random matrix theories describing the
magnetic field dependence of Coulomb blockade peak heights and
the orbital contributions to $g$ factors are the same. 
At present, we do not know of a solution to either puzzle.

One complication in the search for an orbital contribution to 
the $g$ factors measured in Ref.\ \onlinecite{Petta} is that the
main effect of the orbital contribution is to change the average
of the $g$-factor distribution only. Since, for strong
spin-orbit scattering, the average $g$ factor depends on both
the dimensionless spin-orbit coupling $\lambda$ and the
dimensionless orbital contribution $\eta$, cf.\
Eq.\ (\ref{Eq:gsquared}), it is impossible to
characterize what fraction of a measured $g$ factor is the result
of a state's orbital magnetic moment.
The recent development of experimental methods to measure the
entire $g$ tensor\cite{Petta2} opens new avenues to 
investigate the orbital contribution. For weak spin-orbit
scattering, the $g$-tensor distribution depends on the two
parameters $\lambda$ and $\eta$ in a nontrivial way;
even a weak orbital contribution leads to $g$
tensors with, at least, one principal $g$ value larger than
two, see, e.g., Fig.\ \ref{fig:Gweak}. Hence, measurement the
full $g$ tensors for metal grains with weak spin-orbit scattering,
such as large Al grains, eventually doped with a small concentration
of Au,\cite{Salinas}
will allow the independent determination of the orbital contribution.

\acknowledgements

We would like to thank Vinay Ambegaokar, Michael Crawford, Leonid
Glazman, Jason Petta, and Dan Ralph 
for discussions. This work was supported by the Cornell Center for
Materials Research under NSF grant no.\ DMR0079992, by the
NSF under grant 
no.\ DMR 0086509 and by the Packard foundation.



\begin{references}
\bibitem{erratum}
The derivation of the numerical coefficient for the ballistic case in 
Ref.\ \onlinecite{FrahmPichard} contains a mistake. Following  
appendix C of Ref.\ \onlinecite{FrahmPichard}, for a ballistic
sphere with diffuse boundary scattering one has
\begin{eqnarray*}
  \eta^2 &=& {3 \over 4 m^{*2} v_F R^3} 
  \int dt \langle L_z(0) L_z(t) \rangle
  \nonumber \\ &=&
  {m^2 \over 4 m^{*2} R^3}
  \left \langle {|\vec r_0 \times \vec r_1|^2 \over 
  |\vec r_0 - \vec r_1|} \right \rangle,
\end{eqnarray*}
where $L_z$ is the orbital angular momentum in the $z$-direction
and the average is taken over all classical trajectories in the
sphere. Each trajectory is characterized by the points $\vec r_i$
of reflection from the surface; $\vec r_0$ and $\vec r_1$ are the 
points of reflection immediately before
and after time $t=0$. The joint distribution of $\vec r_0$ and
$\vec r_1$ is
$$
  P(\vec r_0,\vec r_1) = {3 \over 16 \pi R^5} 
  |\vec r_0 - \vec r_1|.
$$
The factor $|\vec r_0
- \vec r_1|$, which was not taken into account in Ref.\
\onlinecite{FrahmPichard}, follows from the constraint that
$\vec r_0$ and $\vec r_1$ are last and first boundary points
before and after time $t=0$, instead of an arbitrary pair of boundary
points along the trajectory. (The flight time between $\vec r_0$ and
$\vec r_1$, and hence the probability to be between $\vec r_0$ and 
$\vec r_1$ at $t=0$, is proportional to $|\vec r_0-\vec r_1|$.)
Performing the double integration
over the surface of the sphere then gives Eq.\ (\ref{eq:alpha2}).
\end{references}
\end{document}